\documentclass[a4paper,twoside,fleqn]{article}

\usepackage{espcrc2}
\usepackage{graphicx}

\newcommand{\AmS}{{\protect\the\textfont2 A\kern-.1667em\lower.5ex\hbox{M}\kern-.125emS}}
\newcommand{\bb}{{\beta_1 \over \beta_0}}
\def \Gqqa {\gamma_{qq}^{N(0)}}
\def \Gqga {\gamma_{qg}^{N(0)}}
\def \Ggqa {\gamma_{gq}^{N(0)}}
\def \Ggga {\gamma_{gg}^{N(0)}}
\def \Gqqb {\gamma_{qq}^{N(1)}}
\def \Gqgb {\gamma_{qg}^{N(1)}}
\def \Ggqb {\gamma_{gq}^{N(1)}}
\def \Gggb {\gamma_{gg}^{N(1)}}
\def \Gqqc {\gamma_{qq}^{N(2)}}
\def \Gqgc {\gamma_{qg}^{N(2)}}
\def \Ggqc {\gamma_{gq}^{N(2)}}
\def \Gggc {\gamma_{gg}^{N(2)}}

\def \Cqa {C_{2,q}^{N(1)}} 
\def \Cga {C_{2,g}^{N(1)}} 
\def \Cqb {C_{2,q}^{N(2)}} 
\def \Cgb {C_{2,g}^{N(2)}}

\title{\begin{flushleft}
        {\small DESY 04-218 \hfill {\tt hep-ph/0411110}\\
         October 2004 \hfill SFB/CPP-04-060}
       \end{flushleft}
	Scheme-invariant NNLO evolution for unpolarized DIS structure
        functions}

\author{J.~Bl\"umlein\address[DSZN]{DESY, Platanenallee 6,
        D--15738 Zeuthen, Germany}
        and 
        A.~Guffanti\addressmark[DSZN]}

\begin{document}

\begin{abstract}
\noindent
We discuss the combination of NNLO standard QCD evolution and scheme-invariant
analysis for unpolarized DIS structure functions data as a method to reduce the
theoretical errors on the determination of $\alpha_s(M_Z^2)$ to $\sim 1$\%
in order to match the accuracy forseen for experimental errors from future
high statistics measurements. 
\end{abstract}

\maketitle

\section {INTRODUCTION}
\noindent
The final HERA-II data on unpolarized deeply inelastic scattering (DIS) 
structure functions, combined
with the present world data, will allow to reduce the experimental
error on the strong coupling constant, $\alpha_s(M_Z^2)$, to the level of 
$1$\% 
\cite{ACC}.
On the theoretical side, the next-to-leading order (NLO) analyses have 
intrinsic limitations which 
allow no better than $5$\% accuracy in the determination of $\alpha_s$ 
\cite{NLO}.
In order to match the expected experimental accuracy, analyses of DIS 
structure functions need then to be carried out at the NNLO level, which
requires the knowledge of the 
$\beta$--function and anomalous dimensions at the 3-loop level 
and the massless and massive 2-loop Wilson coefficients. With the recent 
computation of the 3--loop anomalous dimensions
\cite{MVV}, the whole scheme independent set of quantities is known,
thus allowing a complete NNLO study of DIS structure functions.
At the same time we think that combining the standard QCD analysis and 
fits based on scheme-invariant evolution will provide a valuable tool
to reduce theoretical and conceptual uncertainties in high-precision 
analyses aiming at $1$\% accuracy. 

Our final goal is to perform the full NNLO analysis of DIS structure 
functions
aiming at an high-accuracy determination of $\alpha_s$ and the extraction
of a set of parton distribution functions (PDFs) with fully correlated 
errors. 
A complete analysis refers to both the singlet and non-singlet evolution. 
In the present letter we will concentrate 
only on the singlet sector, referring the reader to the recent  
non-singlet analysis  \cite{BBG}.

\section {QCD EVOLUTION EQUATIONS}
\noindent
Evolution equations of DIS structure functions depend, in the standard QCD
approach, on two arbitrary scales which are introduced in the calculation: 
the renormalization and the factorization scales.

The renormalization scale dependence of any observable can be removed only
summing the perturbative series to all orders. Its presence is then 
unavoidable in any fixed order result. Moreover, the dependence of the 
result 
on the variation of this unphysical scale can be used to give a rough 
estimate of the theoretical error due to higher order corrections.

On the other hand, if we consider the dependence of the result on the 
factorization scale we may follow two approaches.
The first one is to consider the evolution of parton distribution functions 
which are process-independent but depend on the adopted factorization scheme. 
The second one is to study evolution equations for physical observables. 
In these equations the r\^ole of anomalous dimensions for mass
factorization is played by physical anomalous dimensions, which are 
independent of the factorization scheme but depend on the process and the 
observables considered.
These two choices define what we call the standard QCD analysis and the 
scheme-invariant analysis.

In the standard QCD analysis one introduces a parameterization for the
different PDFs at a given reference scale. 
The PDFs are then evolved up to the actual scale of the process, 
solving the evolution equations for mass factorization. 
Structure functions are then constructed as a convolution of the PDFs 
and the corresponding Wilson coefficients.
As a last step, a multi-parameter fit is then performed to extract the value 
of the coupling constant and determine the parameters entering the PDF 
parameterization.

In a scheme-invariant analysis the parameterization of the observable 
at the reference scale $Q_0^2$ is extracted from the data. The value of
the observable at the scale $Q^2$ is determined solving the evolution 
equations with physical anomalous dimensions as evolution kernels.
Finally a one-parameter fit is performed to extract the value of 
$\alpha_s(M_Z^2)$.
Once the analysis is completed the parton densities in any factorization 
scheme can be extracted along with the respective experimental errors.
The advantage of considering factorization scheme invariant evolution 
equations for physical observables resides in the fact that the input 
distributions are observables. Full exploitation of this advantage 
requires therefore high statistics measurements to minimize errors on 
the input distributions. Furthermore, the correlations between the 
measured input distributions have to be considered in detail.

Once more we would like to stress that the two analyses are complementary 
and not mutually exclusive. Thus performing both of them and comparing the 
extracted values of $\alpha_s(M^2_Z)$, or correspondingly $\Lambda_{QCD}$, 
provides a test of stability to determine the QCD parameter.

\section {PHYSICAL ANOMALOUS DIMENSIONS}
\noindent
When considering the singlet evolution the quark-singlet and gluon PDFs 
can be mapped into a pair of structure functions via the matrix of Wilson 
coefficients, ${\bf C}^N$, \cite{BRvN}~:
\begin{equation}
\label{map}
\left(
\begin{array}{c}
F_A^N\\
F_B^N
\end{array}\right)=
\left(
\begin{array}{cc}
C_{A,\Sigma}^N & C_{A,g}^N\\
C_{B,\Sigma}^N & C_{B,g}^N
\end{array}\right)
\left(
\begin{array}{c}
\Sigma^N\\
G^N
\end{array}\right)\,.
\end{equation}
In Eq. (\ref{map}), as we will do in the following, we work in Mellin 
space, where convolutions are ordinary products. 

The singlet evolution equations read
\begin{equation}
\frac{d}{dt}
\left(
\begin{array}{c}
F_A^N\\
F_B^N
\end{array}\right)=
-\frac{1}{4}\bf{K}^N
\left(
\begin{array}{c}
F_A^N\\
F_B^N
\end{array}\right)\,,
\end{equation}
where the evolution variable is 
\begin{equation}
t = -\frac{2}{\beta_0}\ln\frac{a_s(Q^2)}{a_s(Q_0^2)}\,.
\end{equation}
The coupling constant $a_s$ is related to the usual strong
interactions coupling via the relation
\begin{equation}
a_s(\mu^2)=\frac{\alpha_s(\mu^2)}{4\pi}
\end{equation}
and its expansion to 3-loops reads
\begin{eqnarray}
a_s(Q^2)&\hspace{-.2cm}=&\hspace{-.3cm}\frac{1}{\beta_0 L}
\left\{1-\frac{\beta_1\ln L}{\beta_0^2 L}\right.\\
&&\hspace{.2cm}\left.
+\frac{\beta_1^2\ln^2 L - \beta_1^2\ln L + \beta_2\beta_0-\beta_1^2}
{\beta_0^4 L^2}\right\},
\nonumber
\end{eqnarray}
where 
\begin{equation}
L=\ln\frac{Q^2}{\Lambda_{QCD}^2}\,,
\end{equation}
the $\beta$ function is defined as
\begin{equation}
\mu^2\frac{d\,a_s(\mu^2)}{d\mu^2}=
-\sum_{n=0}^\infty \beta_n a_s^{n+2}(\mu^2)
\end{equation}
and, in the case of $SU(3)_c$, the coefficients entering up 
to 3-loops are
\begin{eqnarray}
\beta_0 &=& 11 - \frac{2}{3}N_f\,,\nonumber\\ 
\beta_1 &=& 102 - \frac{38}{3}N_f\,,\\
\beta_2 &=& \frac{2857}{2}-\frac{5033}{18}N_f
            +\frac{325}{54}N_f^2\, ,\nonumber
\end{eqnarray}
with $N_f$ the number of flavors.
The physical anomalous dimensions ${\bf K}^N$ can be expressed
in terms of the anomalous dimensions and the Wilson coefficients as 
\cite{BRvN}
\begin{eqnarray}
\label{physad}
K^N_{IJ}&\hspace{-.3cm}=&\hspace{-.3cm}
\Bigg[
-4\frac{\partial C^N_{I,m}(t)}{\partial t}\left(C^N\right)^{-1}_{m,J}(t)
\\
&-&\hspace{-.3cm}
\frac{\beta_0 a_s(Q^2)}{2\beta(a_s\left(Q^2)\right)}
C^N_{I,m}(t)\gamma^N_{mn}(t)\left(C^N\right)^{-1}_{n,J}(t)\Bigg].
\nonumber
\end{eqnarray}
Here $\gamma^N_{mn}$ denotes the unpolarized anomalous dimensions which
are related to the evolution kernels in $x-$space by
\begin{equation}
 \gamma^N_{mn}=-2\int_0^1dx\,x^{N-1}P_ {mn}(x)\,,
 \quad m,n = q,g
\end{equation}
and $C^N_{I,m}$ are the Mellin transforms of the Wilson coefficients 
\begin{equation}
 C^N_{I,m}=\int_0^1dx\,x^{N-1}C_{I,m}(x)\,.
\end{equation}
While the anomalous dimensions and the Wilson coefficients are, 
separately, factorization-scheme dependent quantities, the combinations
(\ref{physad}) defining the physical anomalous dimensions are 
factorization scheme invariants, order by order in 
perturbation theory.

Different pairs of structure function can be taken into consideration, 
in particular:
\begin{itemize}
 \item $F_2$ and $\partial F_2/\partial t$ \cite{FP,BRvN,BB}\,;
 \item $F_2$ and $F_L$ \cite{Cat,BRvN}\,.
\end{itemize}
In the case of polarized DIS a combined $\overline{MS}$ and 
scheme invariant analysis was carried out to NLO for the structure 
functions $g_1$ and $\partial g_1/\partial t$ in \cite{BB}.

Here we present the NNLO physical anomalous dimensions for the coupled 
evolution of the structure functions $F_2$ and $\partial F_2/\partial t$,
alongside with those at LO and NLO, cf.~\cite{BRvN}~:
~\\
{\bf LO~:}
\begin{eqnarray}
K_{22}^{N(0)}&=&0\,,\qquad
K_{2d}^{N(0)}=-4\,,\\
K_{d2}^{N(0)}&=&\frac{1}{4}\Bigg(\Gqqa\Ggga-\Gqga\Ggqa \Bigg)\,,\nonumber\\
K_{dd}^{N(0)}&=&\Gqqa+\Ggga\,.\nonumber
\end{eqnarray}
{\bf NLO~:}
\begin{eqnarray}
K_{22}^{N(1)}&=& K_{2d}^{N(1)} = 0\\
K_{d2}^{N(1)}&=&\frac{1}{4}\Bigg[\Ggga \Gqqb+\Gggb \Gqqa\nonumber\\ 
&-&\Gqgb \Ggqa -\Gqga \Ggqb\Bigg]\nonumber\\
&-&{\beta_1 \over 2 \beta_0} \Bigg(\Gqqa \Ggga-\Ggqa \Gqga\Bigg)\nonumber\\
&+&{\beta_0 \over 2} \Cqa \Bigg(\Gqqa+\Ggga-2 \beta_0\Bigg)\nonumber\\
&-&{\beta_0 \over 2} {\Cga \over \Gqga} \Bigg[(\Gqqa)^2-\Gqqa\Ggga\nonumber\\
&+&2\Gqga\Ggqa-2 \beta_0 \Gqqa\Bigg]\nonumber\\
&-&{\beta_0 \over 2} \Bigg(\Gqqb-{\Gqqa \Gqgb \over \Gqga}\Bigg)\nonumber\\
K_{dd}^{N(1)}&=&\!\Gqqb+\Gggb-\bb \Bigg(\Gqqa+\Ggga \Bigg)\nonumber\\
&-&{2\beta_0\over\Gqga}\Bigg[\Cga\Big(\Gqqa-\Ggga-2\beta_0\Big)\nonumber\\
&-&\Gqgb\Bigg] + 4 \beta_0 \Cqa -2 \beta_1\nonumber
\end{eqnarray}
{\bf NNLO~:}
{\small
\begin{eqnarray}
K_{22}^{N(2)}\hspace{-.2cm}&=&\hspace{-.2cm} K_{2d}^{N(2)} = 0\\
K_{d2}^{N(2)}\hspace{-.2cm}&=&\hspace{-.2cm}\frac{1}{4}\Bigg(\Gqqc\Ggga 
+ \Gqqa\Gggc\nonumber\\
\hspace{-.2cm}&-&\hspace{-.2cm} \Gqgc\Ggqa -\Gqga\Ggqc\nonumber\\
\hspace{-.2cm}&+&\hspace{-.2cm}\Gqqb\Gggb - \Gqgb\Ggqb\Bigg)\nonumber\\
\hspace{-.2cm}&+&\hspace{-.2cm}\frac{\beta_0}{2}\left[\Cqa\left(\Gqqb
+\Gggb\right)\right.\nonumber\\
\hspace{-.2cm}&-&\hspace{-.2cm}\left.\left(\Cqa\right)^2\!\!
\left(\Gqqa+\Ggga\right)
-\!3\Cga\Ggqb\right]\nonumber\\
\hspace{-.2cm}&-&\hspace{-.2cm}\beta_0\left[2\Ggqa\left(\Cgb-\Cga\Cqa\right)\right.
\nonumber\\
\hspace{-.2cm}&-&\hspace{-.2cm}\left.\Cqb\left(\Gqqa+\Ggga\right)
+\Gqqc\right]\nonumber\\
\hspace{-.2cm}&+&\hspace{-.2cm}\beta_0^2\left[3\left(\Cqa\right)^2
-4\Cqb\right]\nonumber\\
\hspace{-.2cm}&+&\hspace{-.2cm}\frac{\beta_1}{2}\left[\Gqqb
+\Cga\Ggqa\right.\nonumber\\
\hspace{-.2cm}&-&\hspace{-.2cm}\left.\Cqa\left(\Gqqa+\Ggga+2\beta_0\right)\right]
\nonumber\\
\hspace{-.2cm}&-&\hspace{-.2cm}\frac{\beta_1}{2\beta_0}\left(\Gqqb\Ggga
+\Gqqa\Gggb\right.\nonumber\\
\hspace{-.2cm}&-&\hspace{-.2cm}\left.\Gqgb\Ggqa-\Gqga\Ggqb\right)\nonumber\\
\hspace{-.2cm}&+&\hspace{-.2cm}\frac{3}{4}\frac{\beta_1^2}{\beta_0^2}
\left(\Gqqa\Ggga-\Gqga\Ggqa\right)\nonumber\\
\hspace{-.2cm}&-&\hspace{-.2cm}\frac{\beta_2}{2\beta_0}
\left(\Gqqa\Ggga+\Gqga\Ggqa\right)\nonumber\\
\hspace{-.2cm}&+&\hspace{-.2cm}\frac{1}{\Gqga}
\left\{2\beta_0^3\Cqa\Cga\right.\nonumber\\
\hspace{-.2cm}&+&\hspace{-.2cm}\frac{\beta_1}{2}\!\Gqqa\!
\left[\Cga\left(\Gqqa-\Ggga\right)-\Gqga\right]\nonumber\\
\hspace{-.2cm}&+&\hspace{-.2cm}\beta_0^2\left[4\Gqqa\left(\Cgb-\Cga\Cqa\right)\right.
\nonumber\\
\hspace{-.2cm}&-&\hspace{-.2cm}\left.\Cga\Cqa\left(\Gqqa-\Ggga
-\frac{\Gqgb}{\Cga}\right)\right.\nonumber\\
\hspace{-.2cm}&+&\hspace{-.2cm}\left.\Cga\left(\Gqqb+\Cga\Ggqa\right)\right]\nonumber\\
\hspace{-.2cm}&+&\hspace{-.2cm}\beta_0\left[\Cga\Cqa\Gqqa\left(\Gqqa-\Ggga\right)\right.
\nonumber\\
\hspace{-.2cm}&+&\hspace{-.2cm}\left.\Gqqa\left(\Cga\Gggb+\Cgb\Ggga\right)\right]
\nonumber\\
\hspace{-.2cm}&+&\hspace{-.2cm}\frac{\beta_0}{2}\left[\Cga\left(\Gqqb\Ggga
+\Ggqa\Gqgb\right.\right.\nonumber\\
\hspace{-.2cm}&-&\hspace{-.2cm}\left.\frac{3}{2}\Gqqa\Gqqb\right)+\Gqqb\Ggqb\nonumber\\
\hspace{-.2cm}&+&\hspace{-.2cm}\left.\left.\left(\Cga\right)^2
\Ggqa\left(\Ggga-\frac{3}{2}\Gqqa\right)\right]\right\}\nonumber\\
\hspace{-.2cm}&+&\hspace{-.2cm}\frac{2\beta_0}{\left(\Gqga\right)^2}\left\{
-\beta_0^2\left(\Cga\right)^2\Gqqa\right.\nonumber\\
\hspace{-.2cm}&+&\hspace{-.2cm}\beta_0\left[-\Cga\Gqqa\Gqgb\right.\nonumber\\
\hspace{-.2cm}&+&\hspace{-.2cm}\left.\left(\Cga\right)^2\Gqqa\left(\Gqqa
-\Ggga
\right.\right.\nonumber\\
\hspace{-.2cm}&+&\hspace{-.2cm}\left.\left.
\frac{\Gqqa\Ggga}{2}\right)\right]
\nonumber\\
\hspace{-.2cm}&-&\hspace{-.2cm}\frac{1}{2}\left[\left(\Cga\right)^2\Gqqa\left(\Gqqa
-\Ggga\right)^2\right.\nonumber\\
\hspace{-.2cm}&-&\hspace{-.2cm}\left.\Gqqa\left(\Gqgb\right)^2\right]\nonumber\\
\hspace{-.2cm}&+&\hspace{-.2cm}\left.\Cga\Gqgb\Gqqa\left(\Gqqa
-\Ggga\right)\right\}\nonumber\\
K_{dd}^{N(2)}\hspace{-.2cm}&=&\hspace{-.2cm}\Gqqc + \Gggc  -4\beta_2\nonumber\\
\hspace{-.2cm}&-&\hspace{-.2cm}4\beta_0\left[\left(\Cqa\right)^2
-2\Cqb\right]\nonumber\\ 
\hspace{-.2cm}&+&\hspace{-.2cm}\left(\frac{\beta_1^2}{\beta_0^2}
-\frac{\beta_2}{\beta_0}\right)
\left(\Gqqa+\Ggga\right)\nonumber\\
\hspace{-.2cm}&-&\hspace{-.2cm}\frac{\beta_1}{\beta_0}\left(\Gqqb+\Gggb
-2\beta_1\right)\nonumber\\
\hspace{-.2cm}&+&\hspace{-.2cm}\frac{4\beta_0}{\Gqga}
\Bigg\{4\beta_0\left(\Cgb-\Cqa\Cga\right)\nonumber\\
\hspace{-.2cm}&+&\hspace{-.2cm}\left(\Gqqa-\Ggga\right)
\left(\Cga\Cqa-\Cgb\right)\nonumber\\
\hspace{-.2cm}&-&\hspace{-.2cm}\Cga\left(\Gqqb-\Gggb-2\beta_1\right)\nonumber\\
\hspace{-.2cm}&-&\hspace{-.2cm}\left(\Cga\right)^2\Ggqa+\Gqgc \Bigg\}\nonumber\\
\hspace{-.2cm}&+&\hspace{-.2cm}\frac{2\beta_0}{\left(\Gqga\right)^2}
\Bigg\{-4\beta_0^2\left(\Cga\right)^2-\left(\Gqgb\right)^2\nonumber\\
\hspace{-.2cm}&-&\hspace{-.2cm}4\beta_0\Cga\Bigg[\Gqgb
-\Cga\left(\Gqqa-\Ggga\right)\Bigg]\nonumber\\
\hspace{-.2cm}&+&\hspace{-.2cm}2\Cga\Gqgb\left(\Gqqa-\Ggga\right)\nonumber\\
\hspace{-.2cm}&-&\hspace{-.2cm}\left[\Cga\left(\Gqqa
-\Ggga\right)\right]^2\Bigg\}\nonumber
\end{eqnarray}}
The analytic structure of the physical anomalous dimensions $K_{ij}^{(n)}$ 
in $x-$space is difficult to obtain since they require the computation 
of the inverse Mellin transform of products of coefficient functions and 
inverse
splitting functions, which are highly non-trivial already in the simplest
cases (for an example see \cite{BK}). We therefore perform the 
evolution in Mellin space. The result, computed for integer $N$ is then
analytically continued to arbitrary complex $N$ using representations
of harmonic sums with arbitrary precision \cite{Harm} and the $x-$space
result is obtained through a single numerical contour integral.

\section {HEAVY FLAVORS CONTRIBUTION}
\noindent
Heavy flavor contributions to DIS structure functions are known to
be sizable in the kinematic region spanned by HERA. An example being
the structure function $F_2$ which, depending on actual 
event kinematics,
can receive contributions from heavy flavors up to the level of
$20-40$\%.
Any analysis aiming at extracting $\alpha_s$ form DIS structure functions 
data with an accuracy of $\sim 1$\% must, therefore, take into account
heavy flavor contributions.
Recently a parameterization of heavy flavor Wilson coefficients in Mellin
space has been derived \cite{AB}, thus allowing a direct incorporation into 
computer codes which solve the evolution equation in Mellin space.

\section {NUMERICAL RESULTS}
\noindent
While full numerical implementation of the NNLO scheme invariant evolution is
almost finished, as a preliminary result, in Fig. 1 and Fig. 2 we present the 
scheme invariant evolution for the structure functions $F_2$ and 
$\partial F_2/\partial t$ at NLO
for four light flavors.
In the present computation the initial form of the observables is built up 
as a convolution of Wilson coefficients and PDFs at the reference scale 
$Q_0^2 = 1 {\rm GeV}^2$, using parameterization of \cite{MRST}. 

\begin{figure}[!t]
\label{f2}
\includegraphics[height=6.7cm]{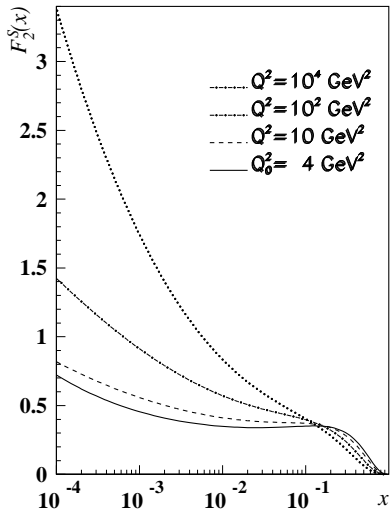}
\vspace{-.8cm}
\caption{NLO scheme invariant evolution for the singlet part of the 
structure function $F_2$ for four light flavors.}
\vspace{-.8cm}
\end{figure}
\begin{figure}[!t]
\includegraphics[height=6.7cm]{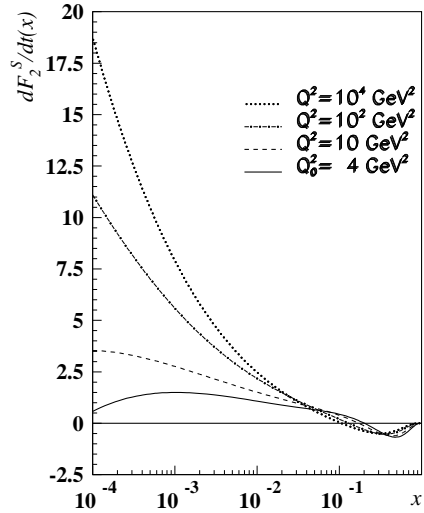}
\vspace{-.8cm}
\caption{NLO scheme invariant evolution for the singlet part of $\partial 
F_2/\partial t$ 
for four light flavors.}
\vspace{-.8cm}
\end{figure}

\section {CONCLUSIONS}
\noindent
The future high precision HERA-II data will allow a reduction 
of  the experimental error on the determination of $\alpha_s$ to $\sim 
1$\%. 
On the theoretical side, the inclusion of NNLO corrections is mandatory to 
cope with such a level of accuracy.
In view of a high accuracy determination of the strong coupling constant
we think that combining the standard $\overline{MS}$ analysis and fits 
based on factorization-scheme invariant evolution could provide a method
to have better control on theoretical and conceptual errors on $\alpha_s$.

\vspace{2mm}\noindent
{\bf Acknowledgment.} This paper was supported in part by DFG 
Sonderforschungsbereich Transregio 9, Computergest\"utzte Theoretische 
Physik.


\begin{thebibliography}{99}
\bibitem{ACC} M. Botje, M. Klein, and C. Pascaud, {\tt hep-ph/9609489}.
\bibitem{NLO} J. Bl\"umlein, S. Riemersma, W.L. van Neerven, and A. Vogt,
              {\tt hep-ph/9609217}. 
\bibitem{MVV}  S.~Moch, J.~A.~S.~Vermaseren and A.~Vogt, 
               Nucl.\ Phys.\ B {\bf B688} (2004) 101; {\bf B691} (2004) 
129.
\bibitem{BBG}  J.~Bl\"umlein, H.~B\"ottcher and A.~Guffanti,
               {\tt hep-ph/0407089}.
\bibitem{BRvN} J.~Bl\"umlein, V.~Ravindran and W.~L.~van Neerven,
               Nucl.\ Phys.\ {\bf B586} (2000) 349.
\bibitem{FP}   W.~Furmanski and R.~Petronzio, 
               Z.\ Phys.\  {\bf C11} (1982) 293.
\bibitem{BB}   J.~Bl\"umlein and H.~B\"ottcher,
               Nucl.\ Phys.\ {\bf B636} (2002) 225.
\bibitem{Cat}  S.~Catani,
               Z.\ Phys.\ {\bf C75} (1997) 665.
\bibitem{BK}   L.~Baulieu and C.~Kounnas,
               Nucl.\ Phys.\ {\bf B155} (1979) 429.
\bibitem{Harm} J.~Bl\"umlein and S.~Kurth, 
               Phys. Rev. {\bf D69} (1999) 014018;\\
	       J.~Bl\"umlein,
	       Comp. Phys. Commun. {\bf 133} (2000) 76;
	       {\bf 159} (2004) 19.
\bibitem{AB}   S.~I.~Alekhin and J.~Bl\"umlein,
               Phys.\ Lett.\ {\bf B594} (2004) 299.
\bibitem{MRST} A.~D.~Martin, R.~G.~Roberts, W.~J.~Stirling 
               and R.~S.~Thorne, Eur.\ Phys.\ J.\ C {\bf 23} (2002) 73

\end{thebibliography}
\end{document}